\documentstyle[pre,aps,multicol,epsf]{revtex}
\begin {document}

\title{ Mean Field  Behavior of Cluster
Dynamics}
\author{
N. Persky\thanks{ Emails: nathanp@vms.huji.ac.il,
radi@opal.co.il, ido@kanter.ph.biu.ac.il,
eytan@elect1.weizmann.ac.il.
}\\
Racah Institute of Physics,
Hebrew University, Jerusalem 91904, Israel \\
R. Ben-Av and I. Kanter \\
Department of Physics, Bar-Ilan University, Ramat-Gan 52100,
Israel \\
E. Domany \\
Department of Physics, Weizmann Institute of Science, Rehovot 76100,
Israel \\
}
\maketitle
%
\begin{abstract}
The dynamic behavior of cluster algorithms is analyzed in the
classical mean field limit. Rigorous analytical results below $T_c$
establish that the dynamic exponent has the value $z_{sw}=1$
for the Swendsen - Wang algorithm
and $z_{uw}=0$ for the Wolff algorithm. An efficient Monte Carlo
implementation is introduced, adapted for using these algorithms
for fully connected graphs. Extensive simulations
both above and below $T_c$ demonstrate scaling and evaluate the finite-size
scaling function by means of a rather impressive collapse of the data.
\end{abstract}
\begin{multicols}{2}
\section{Introduction}

In a series of seminal pioneering papers Fortuin and Kasteleyn \cite{FK}
established an important connection between percolation problems
and the related concept of clusters on the one hand and various
thermodynamic functions of Potts spin models on the other.
These connections allow geometric interpretation of different
properties of these systems, that play a central role in their
critical behavior. For example,
the onset of spontaneous magnetization coincides
with the appearance of an infinite
cluster in the related percolation problem;
the susceptibility
is proportional to the mean cluster size,
and the correlation length
to a typical cluster's radius.

Some years later Swendsen and Wang (SW) \cite{SW} used these ideas in a way
that constituted a breakthrough in a different
sub-field of statistical physics: that of computer simulations of
systems in thermal equilibrium. They
exploited the Fourtuin-Kasteleyn mapping to
define a very efficient Monte Carlo algorithm that
performs large scale non-local moves, by flipping
simultaneously entire clusters of spins.
This dynamics decreases the relaxation time to equilibrium while preserving
detailed balance and ergodicity.
In this way one can overcome (or at least significantly reduce)
critical slowing down at
second order phase transitions \cite{HH}.
"Critical slowing down" is a well known phenomenon;
the relaxation time of a system
diverges
as the critical point is approached.
The manifestation of this {\it real, physical} effect on simulations is that
at criticality
the typical time needed to produce a large set of
decorrelated configurations, that appear with the proper Boltzmann weight,
diverges with the system's size. Hence the standard local
Metropolis-type Monte-Carlo
simulation methods become inefficient.

A wide variety of cluster methods have been applied successfully
to many fields in physics
(second-order phase
transitions, disordered and frustrated systems,
quantum field theories, fermions in a gauge background,
quantum gravity and more \cite{RTK,TBK,Sokal1,binder,Domany2,SokaliC,Sorin,Wolff}). In many cases a
dramatic acceleration of convergence to equilibrium  was
achieved. That is, if the simulation of a system of (linear) size $L$
is performed at the critical point, the relaxation time behaves according
to

\[
\tau \sim L^{z_c}
\]
where $z_c$, the exponent associated with the cluster algorithm is
significantly less than $z_{local}$ of the local methods.

So far only relatively few rigorous results have been derived for
cluster algorithms and the associated relaxation times
and exponents \cite{RTK,Sokal1}.
In this paper we  consider the well known version
of the Ising spin model for which  mean field is the exact solution; that
of $N \rightarrow \infty$ spins, all connected to each other.
By analyzing this system one can calculate the classical,
mean-field value of $z_c$.

The mean-field Ising ferromagnet  has been previously examined analytically
and numerically by Ray, Tamayo and Klein (RTK) \cite{RTK}.
Although the work  presented here agrees with
the conclusions of RTK regarding the SW dynamics, it contains some
improvement.
Firstly, the derivation of the analytical results is improved, employing
the known exact results of Erd\"os et. al.~\cite{erdos} for highly
diluted random graphs, instead of the approximation
(used by RTK) of a Bethe
lattice with large coordination number. We show that this technical improvement
does not  alter the results obtained \cite{RTK} for SW dynamics.
Secondly, we performed simulations,
introducing a new efficient algorithm,
which extend those
of RTK in the following aspects.
RTK looked at the relaxation times $\tau$ of the magnetization autocorrelation
function, measured at a single temperature ($T_c$ of the infinite system)
versus the size of the system. In addition they also studied the temperature
dependence of $\tau$ below $T_c$ at a single value of the system size.
Our  simulations were done on larger systems
using longer running times; in addition, we studied a wide range of system
sizes and temperatures both above and below $T_c$.
Our analysis of the results is based on finite size scaling;
the relaxation times
measured at various temperatures and system sizes are presented using
data collapse, thereby also calculating numerically the
corresponding finite size scaling function. This data collapse of
course takes into account the shift in the transition temperature as a
function of the size of the system. It resolves the
unexpected result which RTK found puzzling,
namely that the slowing down below the transition appeared
to be enhanced in comparison to that at the transition. When the shift
in the transition temperature is properly accounted for, this effect
disappears.

In addition, we present analytical and numerical
results for Wolff (UW) \cite{Wolff} dynamics with the conclusion that
the relaxation times of the autocorrelation function depends on the details of
the global algorithm. However, the upper critical dimension is hypothesized
to be unaffected by the details of the local or the global algorithm.

\section {The Model}

The Hamiltonian  of the fully connected Ising model
 is defined by

\begin{equation}
{\cal H} = -{1\over 2N} \sum_{i \neq j=1}^N S_i S_j \ \  \ \  \ \ S_i = \pm 1.
\label{mfe}
\end{equation}
The statics of this model is solved exactly, using mean-field \cite{exact}.
The result is a
second order phase transition at $\beta_c = 1$, with exponents
$\alpha= 0$, $\nu = 0.5$, $\beta = 0.5$
and $\gamma=1$. The local
(single spin flip) dynamics of this model can also be solved
exactly
using a mean-field approach resulting in (from here we use
$\beta = 1/kT$ unless otherwise stated)
$$
\tau(\beta) \propto (\beta-\beta_c)^{-1}.
$$
Using the fact that $\nu=0.5$ (i.e. $\xi \propto
(\beta-\beta_c)^{-{1\over 2}}$),  it is clear that
$\tau \propto \xi^2$, giving
\[
z_{local} = 2.
\]

The standard SW procedure as implemented for this model is defined as follows.
Let  us assume that the system is in a spin configuration $\{ S_i \}$, $i=1..N$.
In each SW sweep every bond is "activated" (or frozen)
with  probability
\begin{equation}
P_f = 1-exp\{-{\beta \over N} (1+S_iS_j)\}.
\label{pf}
\end{equation}
A bond that is not frozen is called "deleted".
Frozen bonds define
connected clusters, to each of which a spin value ($\pm1$)
is now assigned randomly, giving rise to the new spin configuration. Then
the procedure is repeated. The dynamics of this SW
procedure was first studied for the fully-connected model by
Ray et al \cite{RTK}, who showed that $z_{sw} = 1$.

We present now a different derivation of this result - one
which utilizes exact results from graph theory.

In order to get better insight into the percolation
properties of the fully connected model let us introduce a
few previously known facts about highly diluted graphs \cite{erdos}.
These are relevant to our problem since for $N >>1$ the freezing probability
$P_f$ is very small, and the SW procedure generates precisely the highly
diluted graphs to which we now turn.


Consider a graph of $N$ nodes; on all links connecting pairs of
nodes $i,j$ place
independent
random variables $J_{ij}$, taken from
the probability distribution
\begin{equation}
P(J_{ij})=(1-c/N)\delta(J_{ij})+(c/N) \delta (J_{ij}-1)
\label{m2}
\end{equation}
where the average connectivity $c$
is taken to be ${\cal O}(1)$.

The resulting highly diluted graph has
finite connectivity; in the thermodynamic limit
the probability
that a node has connectivity $k$ (i.e. has non-vanishing link variables
$J_{ij}$
with $k$ other nodes)
follows the Poisson distribution:
$$P(k)=c^k\exp(-c)/k!.$$

Many geometric properties of this diluted graph
are well understood\cite{erdos,KS,Parisi}.
In particular,
the graph undergoes a percolation transition at $c=1$.
That is, the number of nodes that
belong to the largest connected cluster is of order
${\cal O}(\log N)$ for $c <1$,
${\cal O}(N^{2/3})$ at $c=1$, and of ${\cal O}(N)$
for $c>1$. In this percolating regime the
size of the largest cluster is $gN$,
where the parameter $g$ is the solution of the equation \cite{KS}
\[
g=1-e^{-cg}
\]

or
$$ \ln(1-g)=cg.$$
Assuming that for $c=1+\delta $ and  $  \delta \ll 1$ one has $g \ll 1$
we get
$$ cg-g-g^2/2-g^3/3=0 $$
and therefore
$$ 2g^2+3g-6\delta=0 $$
which yields
\begin{equation}
g(1+\delta)= 2\delta-{8\over 3} \delta^2.
\label{del}
\end{equation}

After this aside on random dilute graphs we return to
the SW dynamics for our fully connected Ising spin model.
Since the spins are fully connected, a configuration is fully
parametrized
by $N_+$, the number of positive valued
spins, or
by its magnetization
\[
m=(N_+ - N_-)/N
\]
where $N_-=N-N_+$.
The dynamics is, therefore, completely characterized by specifying the value
of either $N_+$ or $m$ after each Monte Carlo sweep.
Every $SW$ sweep can be thought of as consisting of two steps.
In the first step the spins are divided into two groups, one of positive
spins (all the spins with $S_i = +1$) and the
other negative  (spins
with $S_i = -1$).  Using $P_f$ of (\ref {pf}) we first note that
all the links connecting any positive spin to any
negative one are deleted. The positive (negative) group contains
$$N_{\pm} = {N\over 2}(1 \pm m)$$
spins. Without
loss of generality, we assume that $N_+ > N_-$.

In the second step, each group is examined independently.
The internal links within a group are "activated" with the probability
$P_f = 1-exp(-2\beta/N)$.
 Each group can be thought of now as a random graph
with effective connectivity

\begin{equation}
c_\pm={P_f \times N_\pm}\approx {2\beta N_\pm\over N}.
\end{equation}

Denoting by $t$ the deviation from criticality:
$\beta=1-t$  (recall that  $ \beta_c=1$) we
can write for the larger (positive)
group of spins
\begin{equation}
c_+=(1-t)(1+m).
\end{equation}

We can now view the act of freezing a bond as choosing $J_{ij}=1$ in the
generation of the dilute graph discussed above. For that problem,
as was mentioned previously,  $c=1$ is a critical value.
Therefore
in the disordered phase of our spin model
$t>0$ and $\beta < \beta_c$ so that both $c_+,c_-<1$
and the
SW freeze-delete procedure creates
only small (non-macroscopic) clusters.
On the other hand, at temperatures just below criticality, i.e.
when $t<0$, we have $c_+>1$ but $c_- <1$. Hence only
{\it one} macroscopic cluster (of positive spins),
whose size is proportional to the lattice size, is generated,
while the rest of the
clusters are small.

Flipping of this single macroscopic
cluster is the dominant element of the dynamics.
The rest of the clusters are small, there are ${\cal O}(N)$ of them, and
therefore the extent that their flip influences the magnetization
is negligible
compared to that of the big cluster.

In what follows we derive recursion for $N_+^l$ and $m_l$, the values
that the number of the majority spins and  the magnetization take after
$l$ steps of the SW procedure. We use the notation $N_+$, but mean the
number of spins in the state with majority
(even when they are negative).

The small clusters become $\pm 1$ with probability $1/2$ after a sweep:
neglecting fluctuations of their contribution, and assuming that the single
large spanning cluster was and remains positive, we can write

\begin{equation}
N_+^{l+1}={1\over 2} N_-^l+{1\over 2}[N_+^l-N_+^lg(c_+)]+N_+^lg(c_+)
\end{equation}
\begin{equation}
N_-^{l+1}={1\over 2} N_-^l+{1\over 2}[N_+^l-N_+^lg(c_+)]. \label{nsw}
\end{equation}

Denoting $n_l=N_+^l/N$ equations (7-8) can be rewritten as a recursion for the
magnetization:
\begin{equation}
m_{l+1} =n_{l}g(2 \beta n_{l})={1+m_l \over 2}g[(1-t)(1+m_l)].
\label{ml1}
\end{equation}
For simplicity we keep
$ m $ positive (or, in other words, our recursion is actually for $|m|$).

Near the transition the argument of the function $g$ is near its critical
value and we can identify
 $\delta$  (see Eq. (4)) by
$$\delta =-t+m_l-tm_l. $$

Equation (4) is now given (up to terms of order $t^{3/2}$) by

\begin{eqnarray}
m_{l+1}&=&
{1+m_l \over 2} [2(-t+m_l-tm_l)-{8\over3}(m_l^2-tm_l)] \nonumber \\
&=&m_l-t+{2\over3}m_lt-{1\over3}m_l^2,
\label{ml2}
\end{eqnarray}
which leads to the differential form:
\begin{equation}
{dm \over dl}=-t+{2\over3}m_lt-{1\over3}m_l^2 \label{dif}
\end{equation}
where $l$, the discrete index of the sweep has become a continuous
"time".
For $t<0$ Eq. (\ref{dif}) has a stationary point at
\begin{equation}
m^\star=(-3t)^{1\over 2}.
\end{equation}
Near this point we write
$$m=m^\star+\epsilon,$$
and linearize equation  (\ref{dif}) in $\epsilon$:

\begin{equation}
{d\epsilon \over dl}={2\over3}\epsilon t-
{2\over3}m_l^\star\epsilon=
{2\over3}\epsilon t-{2\over{\sqrt3}}\epsilon (-t)^{1\over 2}.
 \label{dif1}
\end{equation}
For  $|t| \ll 1$ the  $(-t)^{1\over 2} $ term is dominant
so one can conclude that relaxation
is dominated by
$$\epsilon \propto {\exp \left[ -l/\left(
{2\over{\sqrt{3}}}(-t)^{1\over 2} \right) \right]}, $$
and therefore in the ordered phase we find the relaxation time
\begin{equation}
\tau^{sw} = {{\sqrt3}\over 2} (-t)^{-{1\over 2}}. \label{tsw}
\end{equation}
\noindent
Using the definition
$$ \tau = (-t)^{-\nu z} $$
yields our result for the mean field value of the
dynamic exponent of the Swendsen-Wang procedure:
$$ z_{sw}=1. $$


\section{ Numerical Results}

\subsection{The Algorithm}
The standard version \cite{SW} of the SW algorithm visits, during a single
sweep, each bond once and takes a freeze/delete decision~\cite{foot}.

The fully connected system (Eq. \ref{mfe}) contains $N$ spins and
$N^2/2$ links; consequently the
usual SW algorithm requires ${\cal O} (N^2)$ operations for
one sweep of the system.
For our choice of $P_f$ (Eq. \ref{pf}),
however, only a very small fraction of bonds is eventually
frozen: even though
every spin has ${\cal O} (N)$ links,  only ${\cal O} (1)$
are activated.
Moreover, as explained in the Introduction,
in the fully connected model the only relevant
parameter that determines the system's state is its magnetization, or
$N_+$. Hence to get the new configuration {\it after } the SW sweep,
all we have to determine is the new value of $N_+$. To do this we do
not have  to scan all ${\cal O} (N)$
bonds of a certain spin; rather, when the algorithm "visits" a new
spin that belongs to a cluster, it suffice to
use the binomial distribution
to determine the {\bf number} of activated bonds it has (i.e. the
number of new
members of the cluster).

Therefore the SW algorithm can be implemented for the fully connected model
by
manipulating a few {\it numbers}, instead of the values taken by the spins.
This approach simplifies  the algorithm, is much more efficient,
and uses a very small memory.\\

We adopted the spirit of the Wolff algorithm to grow all SW clusters
during a sweep. At a given moment we know the number of spins that
have been assigned to our presently growing cluster. New spins that have
been added to the cluster have to be tested for the number of
frozen bonds connecting them to {\it free} spins (that do not yet belong to
any cluster and have the same sign as the presently growing cluster
and hence are candidates for joining it).
The spins that are known to belong to the cluster
but have not yet been tested are in a {\it "stack"}.
As a cluster is growing, the algorithm tracks and modifies the
following parameters:

\begin{itemize}
\item $N_c= number\  of\  spins\  in\  the\  growing\  cluster$
\item $N_s= number\  of\  spin\  in\  the\  stack$
\item $N_f=number\  of\ "free\ spin"$
\end {itemize}
A SW sweep starts at a
configuration with
$N_+$ "up" spins and $N_-$ "down". Say we start to generate clusters of the
up spins. The following steps are taken:

\begin{enumerate}
\item Assign one spin to the growing cluster: set $N_c=N_s=1$, $N_f=N_+-1$.
\item Test the spin: that is, using the
binomial distribution, determine $N_{add}$,
the number of frozen links our spin will get, connecting it
to presently free spins:
\begin{equation}
{\rm Prob}(N_{add})= \left( \begin{array}{c} N_f \\ N_{add} \end{array} \right)
P_f^{N_{add}} \left( 1-P_f \right)^{N_f-N_{add}}.
\label{eq:binom}
\end{equation}
\item Update :
\begin{itemize}
\item $N_c =N_c+N_{add}$,
\item $N_s=N_s+N_{add}-1$,
\item $N_f=N_f-N_{add}$.
\end {itemize}
\item
Perform another test (as long as $N_s>0$) using step
 $\#2 $.
Repeat $2-4$ until the stack is empty ($N_s=0$). construction of
the cluster has now been completed.

\item
Flip the cluster  with probability ${1\over 2}$.
If the cluster has flipped,  the value of the magnetization changes:
\[
M \rightarrow M-2N_c.
\]

\item  Reset $N_c=1$, $N_s=1$, $N_f=N_f-1$ and repeat $2-6$
to construct new clusters, until
there are no more free spins ($N_f=0$).

\item Go to the second component (of down spins) and apply $1-6$ with $N_-$.
Note that now when the cluster flips the magnetization {\it increases}
by $2N_c$.
\end{enumerate}

Note that
{\bf all} the operations of our algorithm are on a few {\it numbers}, namely
we do not really define spins, just manipulate a few counters that were
defined above.

For the Wolff version one can apply the same algorithm
and grow only the first cluster; the decision of starting it from
an up or down spin as seed is done at random
 (with probability proportional
to the number of up/down spins).

\subsection {Finite Size Scaling Analysis}

Finite size scaling is a standard method \cite{Fisher71,Barber} to extract
 critical indices
from numerical data obtained by simulating finite systems of linear size
$L$. Denote by $t$ the reduced temperature variable, and by  ${\cal B}$ a
thermodynamic
quantity that exhibits singular behavior in the $L \rightarrow
\infty$ limit, characterized by the exponent $\theta$:
\begin{equation}
{\cal B} \propto t ^{-\theta}.
\label{eq:Bt}
\end{equation}
For a finite system clearly ${\cal B}$ is an analytic function of $t$ and
the finite size scaling function exhibits a crossover; when $t$ is such that
the correlation length $\xi \sim t^{-\nu}$ is
much smaller than the system size $L$
the function behaves like~(\ref{eq:Bt}); in the $L \rightarrow \infty$ limit all
dependence on $L$ should disappear. On the other hand,
for $\xi \leq L$ the finite
size effects wipe the singularity out.

The  scaling hypothesis \cite{Fisher71}
asserts that the crossover scaling function depends
on $L$ only through the ratio $ L/ \xi$ and one expects the form

\begin{equation}
 {\cal B} (t,L) =L^{\theta/\nu} f(L/ \xi)=\xi^{\theta/\nu} g(L/ \xi).
\label{eq:OB}
\end{equation}

This scaling form includes implicitly
the finite size shift of $T_c(L)$ which defined by the temperature at which
the measured quantity has a peak.
The size of this shift generally has the following form:

\begin{equation}
|T_c(L)-T_c(\infty)|
 \propto L^{1/\nu}.
\label{peak}
\end{equation}

The specific scaling functions which we
consider in this paper are:\\
For the susceptibility:
\begin{equation}
\chi (t,L) =L^{\gamma/\nu} f(tL^{1/\nu}),
\label{xi}
\end{equation}
and for the relaxation time:
\begin{equation}
\tau(t,L) =L^zf(tL^{1/\nu}).
\label{taus}
\end{equation}

\subsection{Details of the Simulations}
The Monte Carlo simulations were carried out with the
algorithm described above, for systems that contain
$N=10^4-3\times 10^5 $ spins.

The number of MC sweeps per spin was $10^5$.
The first $10^4$ configurations
of the simulation (at list 500 times the
relaxation time itself)
were discarded in order to ensure that the system reached
equilibrium.

The static equilibrium property that
we measured was
$$\chi'= \beta N (<m^2>-<|m|>^2),$$
which is proportional
to the "real" susceptibility $\chi$ \cite{binder}.
The reason for this choice is that
in finite systems, especially with cluster dynamics, the quantity
$<m^2>-<m>^2$ has no peak at all
and the measured value increases monotonically
as T decreases.
The reason is
that $m$ changes sign
frequently and therefore $<m> \sim 0$.
The method
that determines $\chi$ by
measuring the size of the SW clusters \cite{Wolff} also yields no peak.

Fig 1. shows the susceptibility for different system sizes $L$
as a function of temperature.
One can see the large shift of the peak position
for different sizes
of the system.
\begin{figure}
\narrowtext
\epsfxsize= 240pt
\centerline{ \epsffile{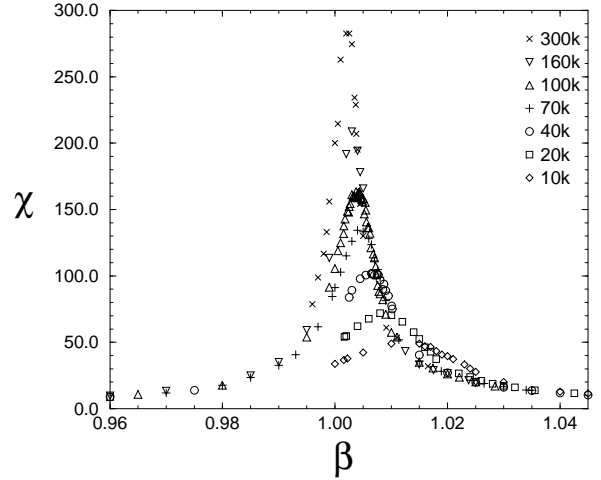}}
\caption{The susceptibility, $\chi$, versus $\beta$ for different
system sizes, measured in the vicinity of the
bulk critical temperature $\beta_c=1$.}
\end{figure}

The most striking manifestation of scaling is {\it data collapse}.
The product  $\chi L^{-\gamma/ \nu}$ should depend only on
$L / \xi \approx L \sqrt{\Delta \beta}$, where $\Delta \beta = \beta -1$.
This
can be seen very clearly in
Fig. 2,  where
the mean field exponents $\gamma/\nu =2$ , $\nu= 1/2$ were used
(we plotted the scaling function versus $(L/\xi)^2$).

\begin{figure}
\narrowtext
\epsfxsize= 240pt
\centerline{ \epsffile{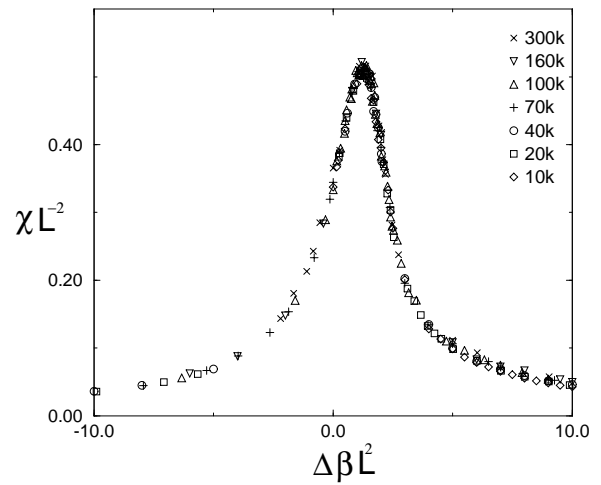}}
\caption{The susceptibility scaling function (Eq. (\ref{xi})).
The mean field value of $\gamma/\nu =2$ was used
for collapsing the data
of Fig. 1.}
\end{figure}

Besides presenting  the scaling function
and verification of the scaling form,
the excellent data collapse that can be seen verifies the accuracy
of our numerical simulations.

The shift  of the effective
$T_c(L)$ is studied in
Fig. 3; indeed one can see that the size of the shift
agrees very well with
Eq. \ref{peak}, with the theoretical mean field value $\nu ={1\over 2}$.

\begin{figure}
\narrowtext
\epsfxsize= 240pt
\centerline{ \epsffile{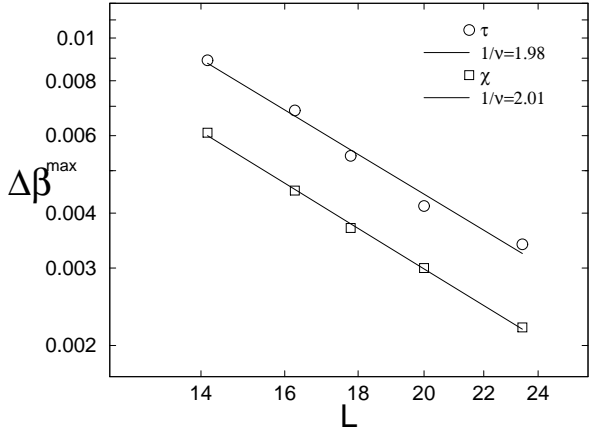}}
\caption{The shift of the peak
$\Delta \beta^{max} =\beta^{^{max}}(L)-1$
of the susceptibility $\chi$,
and of the integrated relaxation time $\tau$,
plotted versus $L=N^{1 \over 4}$.
Non-linear least square fit
yielded the values $1/\nu=1.98\pm0.12$ for $\chi$,
and $1/\nu=2.01\pm0.05$ for $\tau$.}
\end{figure}

The dynamic property that we  measured is the time-dependent
correlation function of the
absolute value of the magnetization.
The measurements were taken after every
sweep.
We considered the normalized
correlation function
of the absolute value of the magnetization
\begin{equation}
\rho(t)={<|m(t')m(t'+t)|>-<|m|>^2> \over{<m^2>-<|m|>^2>}}.
\end{equation}
For which the integrated relaxation time,
defined  by:
\begin{equation}
\tau_{int}={1\over 2} +\sum_{t=1}^\infty{\rho(t)}
\end{equation}
was calculated.

Fig. 4 shows
$\tau$ for different sizes
of the system at various temperatures.
As in the case of the susceptibility, one can
see the large shift of the peak of the scaling
function for different system sizes.
Fig. 3 contains also the scaling analysis of this shift
as a function of $L$: again,
the result agrees very well with
the  theoretical prediction.
Note that although the slope is the same as
in the susceptibility case the shift is much bigger, i.e. the amplitude
is larger.

\begin{figure}
\narrowtext
\epsfxsize= 240pt
\centerline{ \epsffile{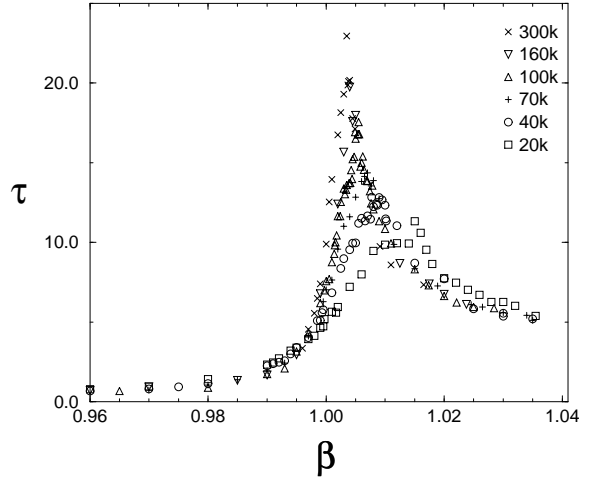}}
\caption{The integrated relaxation time $\tau$,
plotted versus
$\beta$ for different system sizes,
in the vicinity of the
bulk critical temperature $\beta_c=1$.}
\end{figure}

For resonable sizes of the systems (such as
the sizes  that could be treated with
computational power  available to us) one can
not
neglect this shift. For example, calculating $\tau$ (as well as
any other quantity)
for different system sizes {\it at} $T_c(\infty)(=1)$
and trying a fit of the form
$$ \tau(T_c(\infty))=L^z, $$
introduces some systematic errors.

Fig. 5 presents the first measurement (to our knowledge)
of a  scaling function for $\tau$. The theoretically derived value
 $z=1$ was used.

\begin{figure}
\narrowtext
\epsfxsize= 240pt
\centerline{ \epsffile{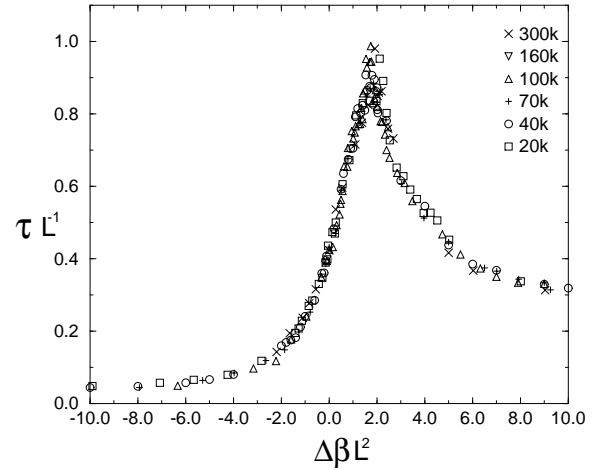}}
\caption{The scaling function for the integrated relaxation time
(Eq. (\ref {taus})).
The value $z=1$ was used
for collapsing the data
of Fig. 4.}
\end{figure}

The evident data collapse  indicates a good fit
to the scaling form presented in Eq. (\ref{taus}) and to the predicted $z$.


\section { Wolff-Dynamics}

Let us start with a remark about SW dynamics.
Going back  to Eq. \ref{nsw}
one can see that for a particular version of SW algorithm,
in which one flips {\bf only} the spanning cluster,
the following equations are obtained

\begin{equation}
N_+^{l+1}=N_+^l-N_+^lg(c_+),
\end{equation}
\begin{equation}
N_-^{l+1}=N_-^l+N_+^lg(c_+). \label{nbc}
\end{equation}
Denoting $N_+^l/N$ by $n_l$, this yields

\begin{eqnarray}
m_{l+1}&=&m_l-2n_{l}g(2 \beta n_{l})\nonumber \\ &=& m_l-(1+m_l)g \biggl[ (1-t)(1+m_l) \biggr] .
\end{eqnarray}
Keeping $m$ positive, one would get instead of Eq. \ref{dif}

\begin{equation}
{dm \over dl}=-2t+{4\over3}m_lt-{2\over3}m_l^2 \label{dif2}.
\end{equation}

Going over the previous calculation (Eqs. \ref{dif}-\ref{tsw})
one gets the same fixed point,
the critical exponents do not change,
and only a factor ${1\over 2}$ multiplies
the prefactor in the expression for $\tau$.
That is,
$$\tau^{sw1}={1\over 2} \tau^{sw}.$$
This result shows that flipping the spanning cluster constitutes the central
component the relaxation process.
Moreover, it shows that  flipping small clusters
together with the flip of the spanning
cluster does not help the relaxation at all!

We turn now to UW dynamics and calculate $\tau^{uw}(\beta)$
in the thermodynamic limit, for $T>T_c$.
We calculate $\tau^{uw}$ in two steps. In the first step we calculate
$\tau^{uw}_c$, the relaxation time measured using every single
cluster update as one unit of time.
In the second step, we normalize these $UW$ time units to regular
time units
($N$ spins are accessed or updated in a single regular unit of time).
To achieve this normalization we
multiply the time measured in $UW$ time units by the ratio of
the average cluster size $<C>$ to the system size $N$.

To calculate $\tau^{uw}_c$ note that
\begin{eqnarray}
\tau^{uw}_c &=& \tau^{sw1} \times {N \over \hat C} =
{1\over 2} \tau^{sw} \times {N \over \hat C} \nonumber \\
&=&{1\over 2} \tau^{sw} \times
{1 \over g} \times {N \over N_+}={\tau^{sw}\over (1+m) g}
\approx {\tau^{sw}\over g},
\label{eq3}
\end{eqnarray}
where $\hat C=gN$ is the size of the macroscopic cluster.
To understand the leftmost equality note
that every cluster
update of the UW procedure has a chance of $\hat C \over N$ to hit the
macroscopic
cluster. The rest of the cluster updates are irrelevant in the thermodynamics
 limit.
When  normalized  to regular units of whole lattice update we get

\begin{equation}
\label{eq4}
\tau^{uw} = \tau^{uw}_c \times {<C> \over N}
\end{equation}
where $<C>$ is the average cluster size in the UW procedure.

We turn now to calculate $<C>$, using the theory of
random graphs   \cite{erdos}.
For the $N_-$ part $c<1$, and the cluster size distribution has
only "small-clusters" with average cluster size $r_1={\cal O}(1)$.
For the $N_+$ part $c>1$ and the cluster size distribution has two components.
The first component is that of
"small-clusters" with average
cluster size $r_2={\cal O}(1)$ and the second component is a single
spanning cluster, whose size depends linearly upon $N_+$ with size
$g(c_+)N_+$. Therefore:
$${<C>\over N} = {1\over N} {\Biggl[}{N_-\over N }r_1
+{N_+\over N }\{\sum_{w=cluster\ size \in N_+} P(w)\times w\}{\Biggr]} $$
$$ \approx
{1\over N}{\Biggl[}{N_-\over N }r_1 +{N_+\over N }\{(1-g(c_+)) r_2 + g(c_+)(
g(c_+)N_+)\}{\Biggr]}. $$
In the limit $N\rightarrow \infty$, the first two terms vanish, and the
leading term is  $\left({N_+\over N}\right)^2\times g^2$.
Thus, we arrive at :

\begin{equation}
{<C>\over N} \approx {(1+m)^2 g^2\over 4} \approx { g^2\over 4}.
\end{equation}
Combining equations (\ref{tsw}),(\ref{eq3}) and (\ref{eq4}), we find that

\begin{equation}
\tau^{uw} \approx \tau^{sw} g(c_+)(1+m)/4 = \tau^{sw}
 g\biggl[(1-t)(1+m)\biggr](1+m)/4,
\end{equation}
Using equation (\ref{del}) we expand $g$ to obtain:
\begin{eqnarray}
\tau^{uw} &\approx&
\tau^{sw} 2(-t+m-tm)(1+m)/4 \approx \tau^{sw}m/2 \nonumber \\
&\approx&
{{\sqrt3}\over 2}(-t)^{-{1\over 2}} \times  \sqrt{3}(-t)^{1\over 2} /2
\approx 0.75 = {\rm Const.}
\label{075}
\end{eqnarray}
Thus we have shown that $\tau^{uw}$ does not depend on $t$ and
hence $z=0$ for $T<T_c$.

\section{ Numerical Results-Wolff}

The Monte Carlo simulations were carried out using the
algorithm described in Sec 3, growing only the first cluster.
Systems that contain
$N=2 \times 10^4-3\times 10^5 $ spins were simulated.
The number of cluster flips was $2 \times 10^6 $ for each run.
The first $10^5$ configurations
of each  simulation
were discarded in order to ensure
equilibration.
As in the SW case, the dynamical property  measured was the
correlation function of the
absolute value of the time-series of the magnetization.
The measurements were done after every
4 cluster updates.
>From the  correlation function we calculated the integrated relaxation time.
Finally we normalized by the average cluster size
as was measured during the simulation.

Fig. 6 shows
$\tau$ for different sizes
of the system at various temperatures.
In order to get the scaling function we performed
data collapse for $\tau$ with
the theoretical value of $z=0$.

\begin{figure}
\narrowtext
\epsfxsize= 240pt
\centerline{ \epsffile{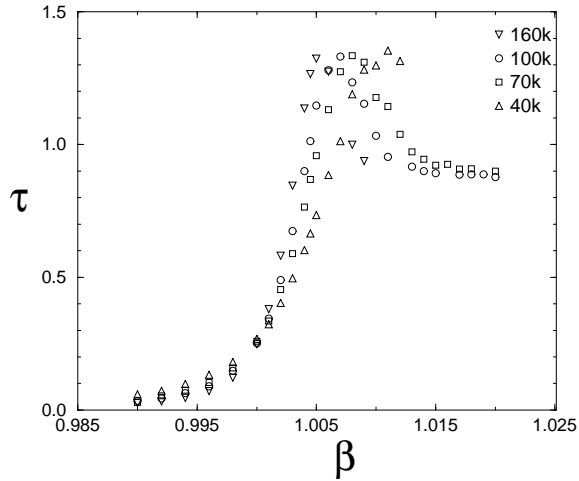}}
\caption
{The Wolff integrated relaxation time $\tau^{uw}$,
plotted versus
$\beta$ for different system sizes,
in the vicinity of the
bulk critical temperature $\beta_c=1$.}
\end{figure}

\begin{figure}
\narrowtext
\epsfxsize= 240pt
\centerline{ \epsffile{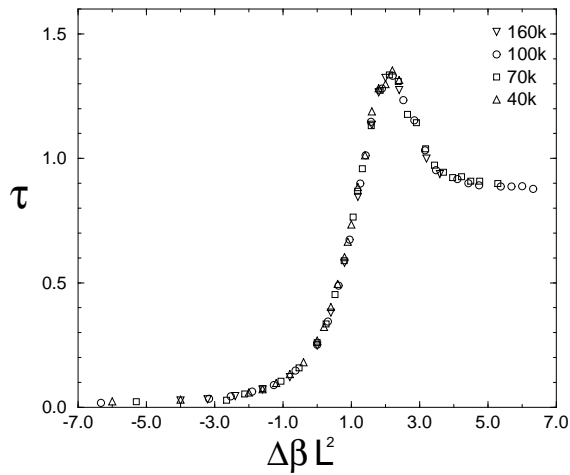}}
\caption{The scaling function for the integrated relaxation time  Eq. (\ref{taus}).
The value $z=0$ was used
for collapsing the data
of Fig. 6.}
\end{figure}

The results is shown in
Fig. 7, in which one can see a very good agreement
with the scaling form presented at Eq. \ref{taus}.
In addition the asymptotic behavior
is consistent with the result of Eq. (\ref{075}), valid for $\beta > \beta_c$
and $L >> 1$ .

\section {Summary and Conclusions}
Cluster algorithms constitute a significant advance in computational
physics. Their associated dynamics is characterized by critical
exponents that differ from those of standard methods. Very few exact
results have so far been derived regarding these exponents.

In the present work we addressed the issue of the
classical limit for the dynamic exponents
of two cluster algorithms. For the Swendsen-Wang algorithm we
found $z=1$; our analytical derivation of this
result is based
on rigorous results from the theory of random graphs. Since the
proof applies only for the low-temperature phase, we measured
the relaxation time $\tau^{sw}$,
using a novel implementation of the SW algorithm
adapted for fully-connected graphs, {\it both above and below} $T_c$.
We established scaling behavior of $\tau^{sw}(t,L)$ by demonstrating
rather conclusive data collapse. Previous work used a Bethe-lattice
approximation for the analytic derivation, and did not present the
full scaling function for the relaxation time.

For the single-cluster algorithm of Wolff we have shown that $z=0$,
and presented numerical evidence, again obtained both above and below $T_c$.
The scaling function was found numerically and data collapse was demonstrated.

\section* {Acknowledgements}
We thank D. Kandel and S. Solomon for discussions, and
R. Swendsen for some most illuminating comments and suggestions.
This research was partially supported by grants from the US-Israel BSF,
the Germany-Israel Foundation (GIF) and
from the Israeli Academy of Sciences.


\end{multicols}

\begin{thebibliography}{99}
\bibitem{foot}
Actually, if both spins linked by a bond have already been assigned to the
same cluster, there is no need to visit the bond.
\end{thebibliography}
\end{document}